\documentclass[journal]{IEEEtran}

\usepackage{amsmath,amssymb,bm}
\usepackage{graphicx}
\usepackage{cite}
\usepackage{booktabs}
\usepackage{algorithm}
\usepackage{algorithmic}
\usepackage{balance}
\usepackage{url}
\usepackage{xcolor}
\usepackage{tikz}
\usetikzlibrary{calc}
\colorlet{red}{black}

\newcommand{\tr}{\mathrm{tr}}
\newcommand{\R}{\mathbb{R}}
\newcommand{\C}{\mathbb{C}}
\newcommand{\re}{\operatorname{Re}}
\newcommand{\Jmat}{\bm{J}}
\newcommand{\wvec}{\bm w}
\newcommand{\avec}{\bm a}
\newcommand{\pvec}{\bm p}
\newcommand{\etavec}{\bm{\eta}}
\newcommand{\Rmat}{\bm R}
\newcommand{\rev}[1]{#1}

\begin{document}

\title{Cram\'{e}r-Rao Bound Optimization for Near-Field ISAC with Extended Targets}

\author{Zongyao Zhao, Zhaolin Wang, Lincong Han, Liang Xu, Jing Jin, \\ Yuanwei Liu, \IEEEmembership{Fellow, IEEE}, Kaibin Huang, \IEEEmembership{Fellow, IEEE}
        % <-this % stops a space

\thanks{Z. Zhao, Z. Wang, Y. Liu, and K. Huang are with \rev{the Department of Electrical and Computer Engineering,} The University of Hong Kong, Hong Kong (\rev{e-mails}: zongyao@hku.hk, zhaolin.wang@hku.hk, yuanwei@hku.hk, huangkb@hku.hk).\emph{(Corresponding authors: Yuanwei Liu and Kaibin Huang)}}

\thanks{L. Han and J. Jin are with Future Research Laboratory, China Mobile Research Institute, Beijing. L. Han is also with China Mobile (Hong Kong) Innovation Research Institute, Hong Kong (\rev{e-mails}: hanlincong@chinamobile.com, jinjing@chinamobile.com).}

\thanks{L. Xu is with China Mobile Hong Kong Company Limited, Hong Kong (\rev{e-mail}: xu.bright@gmail.com).}

}

\maketitle

% ============================================================
\begin{abstract}
Near-field integrated sensing and communication (ISAC) requires target models beyond the point-target abstraction when the target has a non-negligible spatial extent. \rev{In this letter, a geometry-aware transmit design is developed for a parametric extended target (ET) described by its center, orientation, and size under spherical-wave propagation. The CRB for the geometric parameters is formulated around a nominal ET state, an exact ET-aware reduced subspace is identified for the lifted covariance formulation, and a reduced-dimensional semidefinite relaxation (SDR) is developed under signal-to-interference-plus-noise ratio (SINR) and power constraints. Simulation results show lower CRB values than point-target and geometry-agnostic baselines together with substantially reduced runtime for large arrays.}
\end{abstract}

\begin{IEEEkeywords}
Integrated sensing and communication, near-field, extended target, Cram\'{e}r-Rao bound, beamforming, semidefinite relaxation, covariance optimization.
\end{IEEEkeywords}

% ============================================================
\section{Introduction}\label{sec:intro}

Integrated sensing and communication (ISAC) is expected to be a key capability of sixth-generation (6G) wireless systems~\cite{liu2022isac_survey}. With large-aperture arrays, many ISAC links already operate in the near field, where spherical-wave propagation couples angle and range, changing both beam design and parameter estimation~\cite{cui2022nf_channel,li2024nfispac}. In this regime, a spatially extended object is no longer represented well by a single scatterer. Different parts of the object induce different phases, ranges, and geometric sensitivities across the array. The transmit covariance must therefore satisfy user signal-to-interference-plus-noise ratio (SINR) constraints while probing a parametric extended target (ET).

Existing near-field ISAC beam design still mainly follows two lines: communication-oriented beamfocusing~\cite{wang2025beamfocusing} and sensing-oriented beamforming based on point-target CRB formulations, robust localization surrogates, or multi-target detection~\cite{liu2021crb_jrc,zhao2024robustcrb,zhao2025multitarget}. For extended targets, CRB designs have been studied mostly in far-field or conventional ISAC settings, including CRB-rate tradeoff formulations, geometric ET beam design, and three-dimensional ET models~\cite{hua2022et_tradeoff,wang2024twc_et,wang2024globecom_3det}. Near-field ET studies instead emphasize range-only bounds and electromagnetic propagation models~\cite{thiran2024spawc,chen2024jsac_em,zhao2026etcl}. Structural beamforming subspace results are also available for point-target CRB design~\cite{fang2025obs} and non-parametric ET response estimation~\cite{zhao2026lowdim}, but not for parametric near-field ETs with center, orientation, and size estimation. What remains missing is communication-constrained covariance design that captures the geometry Jacobian directions of a parametric ET contour.

Here we take a different route. The CRB for the geometric parameters is formulated around a nominal operating point with a calibrated scattering profile, so the dependence on the transmit covariance stays linear while the joint sensitivity of center, orientation, and size is retained. We then show that the lifted covariance problem admits an exact ET-aware reduced subspace spanned by the user channels and ET steering/Jacobian directions, which leads to a reduced-dimensional semidefinite relaxation (SDR) under SINR and power constraints. The main contributions are as follows.

\begin{itemize}\setlength{\itemsep}{0pt}\setlength{\topsep}{1pt}
  \item \rev{We propose a novel geometry-aware CRB characterization for parametric near-field ET beamforming design that remains linear in the transmit covariance.}
  \item \rev{We identify an exact ET-aware reduced-subspace method that yields an equivalent reduced-dimensional SDR without loss of relaxed optimality.}
  \item \rev{Simulation results show that the proposed design achieves sensing gains over focus, point-target, and geometry-agnostic baselines together with large-array runtime savings.}
\end{itemize}

\noindent\textit{Notation:} Bold lower- and upper-case letters denote vectors and matrices. $(\cdot)^{\mathsf T}$ and $(\cdot)^{\mathsf H}$ denote transpose and Hermitian transpose; $\tr(\cdot)$, $\re\{\cdot\}$, $\mathcal{R}(\bm A)$, and $\operatorname{span}(\cdot)$ denote the trace, real part, column space, and linear span, respectively. $\mathcal{CN}(\bm \mu,\bm \Sigma)$ denotes the circularly symmetric complex Gaussian distribution with mean~$\bm \mu$ and covariance~$\bm \Sigma$. $\bm I$ denotes the identity matrix with appropriate dimension. $\bm A\succeq0$ means that $\bm A$ is positive semidefinite, and $\|\cdot\|$ denotes the Euclidean norm.

% ============================================================
\section{\rev{System Model and Problem Formulation}}\label{sec:system}

%\subsection{\rev{Narrowband ISAC System Setup}}

We consider a narrowband monostatic near-field ISAC system in which an $N$-antenna base station (BS) serves $K$ single-antenna communication users while sensing one ET. The BS employs a uniform linear array (ULA) with half-wavelength spacing $d=\lambda/2$ at carrier frequency~$f_c$. Both the users and the ET are assumed to lie in the near-field region, i.e., within the Fresnel distance $R_F=2D^2/\lambda$, where $D=(N-1)d$ denotes the array aperture.

\subsection{Communication Model}

\rev{To fully exploit the spatial degrees of freedom (DoFs) for both communication and sensing, we consider the following composite ISAC signal transmitted by the BS:}
\begin{equation}\label{eq:tx_signal}
  {\color{red}\bm x(t) = \sum_{k=1}^{K} \wvec_k c_k(t) + \bm s_0(t),}
\end{equation}
where \rev{$\wvec_k\in\C^N$ and $c_k(t)$ denote the beamformer and data symbol for user~$k$, while $\bm s_0(t)$ denotes a dedicated sensing component with second-order moment $\mathbb{E}[\bm s_0(t)\bm s_0^{\mathsf H}(t)]=\bm W_0\succeq0$, independent of $\{c_k(t)\}$. The data symbols satisfy $\mathbb{E}[|c_k(t)|^2]=1$ and are mutually independent. Under these assumptions, the transmit covariance can be written as}
\begin{equation}
  {\color{red}\Rmat_x = \mathbb{E}[\bm x(t)\bm x^{\mathsf H}(t)] = \sum_{k=1}^{K} \wvec_k \wvec_k^{\mathsf H} + \bm W_0.}
\end{equation}

The received signal at user~$k$ is given by
\begin{equation}
  {\color{red}y_k(t) = \bm h_k^{\mathsf H} \bm x(t) + z_k(t),}
\end{equation}
where \rev{$\bm h_k=\alpha_k\avec(\pvec_k)\in\C^N$ is the near-field channel for user~$k$, $\pvec_k$ denotes the user location corresponding to range~$r_k$ and angle~$\theta_k$, and $z_k(t)\sim\mathcal{CN}(0,\sigma_c^2)$ is additive white Gaussian noise. For any point location $\pvec\in\R^2$, the near-field steering vector is defined as}
\begin{equation}
  {\color{red}[\avec(\pvec)]_n = \frac{1}{\sqrt{N}} e^{-j\frac{2\pi}{\lambda}(d_n(\pvec)-d_{\mathrm{ref}}(\pvec))},}
\end{equation}
\rev{where $d_n(\pvec)=\|\pvec-\bm q_n\|$ is the distance from~$\pvec$ to the $n$th array element, $d_{\mathrm{ref}}(\pvec)=\|\pvec\|$ is the reference distance, and $\bm q_n$ denotes the position of the $n$th array element. Thus, $\avec(r_k,\theta_k)$ used in the user channel is simply shorthand for $\avec(\pvec_k)$.} The SINR of user~$k$ is
\begin{equation}\label{eq:sinr}
  \gamma_k = \frac{|\bm h_k^{\mathsf H} \wvec_k|^2}{\sum_{j \neq k} |\bm h_k^{\mathsf H} \wvec_j|^2 + \bm h_k^{\mathsf H}\bm W_0 \bm h_k + \sigma_c^2}.
\end{equation}

\subsection{\rev{Sensing Model for Extended Targets}}

\begin{figure}[!t]
\centering
\definecolor{etred}{RGB}{190,30,30}
\begin{tikzpicture}[>=stealth, font=\footnotesize, scale=0.5]

  % ===== ULA =====
  \foreach \i in {0,...,12}{
    \pgfmathsetmacro{\xp}{-1.8+\i*0.3}
    \fill[black] (\xp, 0) circle (1.8pt);
  }
  \draw[thick, black] (-2.1, 0) -- (1.9, 0);
  \node[below=3pt, black] at (0, 0) {BS/ULA};

  % ===== Near-field wavefronts =====
  \foreach \r in {1.8, 3.0, 4.2}{
    \draw[black!15, thin] (0,0.1) ++(90-25:\r) arc[start angle={90-25}, end angle={90+25}, radius=\r];
  }

  % ===== Extended Target (ellipse) =====
  \def\phiA{30}
  \coordinate (C) at (0.2, 5.2);
  \def\sA{2.2}
  \def\sB{0.68}

  \begin{scope}[shift={(C)}, rotate=\phiA]
    % Outer ellipse
    \draw[thick, etred!70, dashed, fill=etred!4]
      (0,0) ellipse [x radius=\sA, y radius=\sB];
    % Outer scattering points (non-uniform)
    \foreach \ang in {5,30,48,80,110,145,165,190,220,240,275,305,330,350}{
      \fill[etred!90!black] ({\sA*cos(\ang)}, {\sB*sin(\ang)}) circle (2.5pt);
    }
    % Interior scattering points (scattered freely)
    \foreach \p in {(0.7,0.18),(-0.4,0.22),(-0.9,-0.08),(0.25,-0.2),(1.1,-0.1),(-0.15,0.0),(0.5,-0.3)}{
      \fill[etred!70!black] \p circle (2pt);
    }
  \end{scope}

  % Center point
  \fill[etred] (C) circle (2.5pt);

  % --- Only p_m(eta) label ---
  \pgfmathsetmacro{\ptx}{0.2 + cos(\phiA)*\sA*cos(20) - sin(\phiA)*\sB*sin(20)}
  \pgfmathsetmacro{\pty}{5.2 + sin(\phiA)*\sA*cos(20) + cos(\phiA)*\sB*sin(20)}
  \node[black, anchor=south west] at (\ptx+0.25, \pty+0.35) {$\pvec_m(\etavec)$};
  \draw[->, thin, black] (\ptx+0.22, \pty+0.28) -- (\ptx+0.05, \pty+0.07);

  % ===== Users =====
  \coordinate (U1) at (-3.8, 2.5);
  \draw[thick, black, fill=white, rounded corners=1.5pt]
    ($(U1)+(-0.22,-0.3)$) rectangle ++(0.44, 0.6);
  \draw[black!50] ($(U1)+(-0.13,-0.17)$) rectangle ++(0.26, 0.15);
  \fill[black!50] ($(U1)+(0, 0.18)$) circle (0.6pt);
  \node[left=4pt, black] at ($(U1)+(-0.22,0)$) {User $1$};

  \coordinate (U2) at (4.0, 3.2);
  \draw[thick, black, fill=white, rounded corners=1.5pt]
    ($(U2)+(-0.22,-0.3)$) rectangle ++(0.44, 0.6);
  \draw[black!50] ($(U2)+(-0.13,-0.17)$) rectangle ++(0.26, 0.15);
  \fill[black!50] ($(U2)+(0, 0.18)$) circle (0.6pt);
  \node[right=4pt, black] at ($(U2)+(0.22,0)$) {User $2$};

  % ===== Signal paths =====
  \draw[densely dashed, black, ->, shorten >=3pt]
    (-0.15, 0.15) -- ($(U1)+(0.06,-0.3)$);
  \draw[densely dashed, black, ->, shorten >=3pt]
    (0.15, 0.15) -- ($(U2)+(-0.06,-0.3)$);
  \draw[dotted, thick, etred!70, ->, shorten >=5pt]
    (0, 0.15) -- ($(C)+(0,-0.35)$);

  % Link labels
  \node[black, rotate=52] at (-2.2, 1.5) {\scriptsize comm.};
  \node[etred!80!black, right] at (0.15, 2.5) {\scriptsize sensing};

\end{tikzpicture}
\caption{\rev{Near-field ISAC system with a parametric ET and representative scattering points $\pvec_m(\etavec)$ under spherical-wave propagation.}}
\label{fig:et_model}
\end{figure}
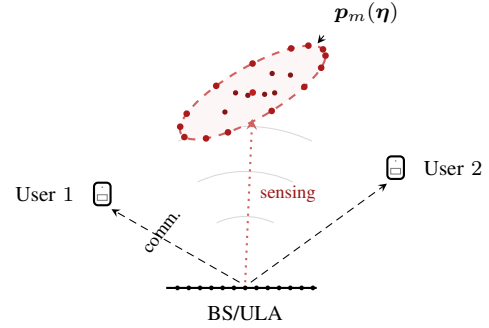

The ET \rev{can be described} by the geometric parameter vector
\begin{equation}
  \etavec \in \R^{D_\eta},
\end{equation}
where \rev{the entries of $\etavec$ specify the adopted target contour. More generally, the extended target is represented by $M$ representative scattering points whose positions are parameterized by the shared geometric vector through a differentiable mapping $\pvec_m(\etavec)\in\R^2$, $m=1,\ldots,M$. This low-dimensional geometric parameterization captures the dominant center, orientation, and extent of an extended target without introducing pointwise position parameters for all scattering points; Fig.~\ref{fig:et_model} illustrates the resulting ET geometry and representative scattering points. The subsequent CRB derivation only requires the local Jacobians $\bm D_m=\partial \pvec_m/\partial\etavec$ to be well defined. Section~IV adopts an elliptical instance as a representative smooth-contour model.} A co-located monostatic array is assumed, and each representative scattering point is modeled as an isotropic reflector with coefficient~$\beta_m$. For tractability, the scattering coefficients are treated as known, e.g., obtained from a prior calibration or tracking stage, and held fixed over the coherence interval. Under the monostatic assumption, the transmit and receive steering vectors coincide, and \rev{the steering vector of the $m$th scattering point is simply $\avec_m\triangleq \avec(\pvec_m)$, following the same phase-exact near-field spherical-wave model defined above.}

Under the above model, the echo observed at the BS \rev{can be written as}
\begin{equation}\label{eq:echo}
  {\color{red}\bm y_s(t) = \sum_{m=1}^{M} \beta_m (\avec_m^{\mathsf H} \bm x(t)) \avec_m + \bm n_s(t) = \bm G \bm x(t) + \bm n_s(t),}
\end{equation}
where \rev{$\bm G = \sum_m \beta_m \avec_m \avec_m^{\mathsf H}$ is the ET response matrix and $\bm n_s(t)\sim\mathcal{CN}(\bm 0,\sigma_s^2 \bm I_N)$. The sensing objective is to estimate the geometric vector~$\etavec$ from the echo snapshots $\{\bm y_s(t)\}$ with the transmitted signals $\{\bm x(t)\}$ known at the BS.}

% ============================================================
\subsection{\rev{Geometry-Aware CRB for Extended Targets}}\label{sec:crb}

\rev{For CRB analysis, we consider $T$ sensing snapshots and work in the covariance domain with $\frac{1}{T}\sum_{t=1}^{T}\bm x_t\bm x_t^{\mathsf H}=\Rmat_x$, so the FIM depends on the data only through $\Rmat_x$ and the factor~$T$.}

With the scattering profile~$\bm{\beta}=[\beta_1,\ldots,\beta_M]^{\mathsf T}$, the CRB is formed directly for the geometric vector~$\etavec$. For snapshot~$t$, the conditional mean of the received echo is
\begin{equation}
  \bm{\mu}_t(\etavec) = \mathbb{E}[\bm y_{s,t}\mid \bm x_t] = \bm G(\etavec,\bm{\beta})\bm x_t.
\end{equation}
Since the noise covariance is $\sigma_s^2\bm I_N$, the deterministic Fisher information matrix (FIM) of~$\etavec$ has entries
\begin{equation}
  [\bm J(\etavec)]_{pq}
  = \frac{2}{\sigma_s^2}\sum_{t=1}^{T}
  \re\!\left\{
    \left(\frac{\partial \bm{\mu}_t}{\partial \eta_p}\right)^{\mathsf H}
    \left(\frac{\partial \bm{\mu}_t}{\partial \eta_q}\right)
  \right\}.
\end{equation}
Define
\begin{equation}
\begin{aligned}
\bm G_m &= [\partial \avec(\pvec)/\partial p_x,\, \partial \avec(\pvec)/\partial p_y]_{\pvec=\pvec_m}\in\C^{N\times 2},\\
\bm D_m &= \partial \pvec_m/\partial \etavec\in\R^{2\times D_\eta},\\
\bar{\bm A}_m &= \bm G_m\bm D_m\in\C^{N\times D_\eta}.
\end{aligned}
\end{equation}
The partial derivative of the received mean with respect to the $p$-th geometric parameter is
\begin{equation}
  \frac{\partial \bm{\mu}_t}{\partial \eta_p} = \bm F_p \bm x_t,
\end{equation}
where
\begin{equation}
  \bm F_p = \sum_{m=1}^M \beta_m \left(\avec_m [\bar{\bm A}_m]_{:,p}^{\mathsf H} + [\bar{\bm A}_m]_{:,p} \avec_m^{\mathsf H}\right).
\end{equation}
Substituting the derivatives above and using the covariance relation yields the closed-form expression
\begin{equation}\label{eq:fim_quad}
  [\Jmat(\etavec;\Rmat_x)]_{pq}
  = \frac{2T}{\sigma_s^2}\,\tr(\bm Q_{pq}\Rmat_x),
\end{equation}
where
\begin{equation}\label{eq:q_def}
  \bm Q_{pq} = \frac{1}{2}\left(\bm F_p^{\mathsf H} \bm F_q + \bm F_q^{\mathsf H} \bm F_p\right)
\end{equation}
are precomputable Hermitian matrices.
\rev{Accordingly, the associated CRB matrix and the scalar metric $\mathrm{CRB}_{\eta}$ are defined as}
\begin{equation}
{\color{red}
\begin{aligned}
\bm C_{\eta}(\etavec;\Rmat_x)&=\Jmat(\etavec;\Rmat_x)^{-1},\\
\mathrm{CRB}_{\eta}(\etavec;\Rmat_x)&=\tr\!\bigl(\bm C_{\eta}(\etavec;\Rmat_x)\bigr).
\end{aligned}
}
\end{equation}
Thus, the CRB for the geometric parameters depends on the target geometry~$\etavec$, the calibrated scattering profile~$\bm{\beta}$, and the transmit covariance~$\Rmat_x$ through the matrices~$\bm Q_{pq}$. In particular, the ET geometry enters through the steering and Jacobian directions collected in~$\{\bm Q_{pq}\}$, whereas the beam design itself remains a covariance optimization. The linear dependence of~$\Jmat(\etavec;\Rmat_x)$ on~$\Rmat_x$ is the key property enabling the subsequent convex reformulation. Note that $\Jmat$ is positive definite whenever $\Rmat_x$ excites at least $D_\eta$ linearly independent directions spanned by the columns of $\{\bm F_p\}$.

% ============================================================
\subsection{\rev{Problem Formulation}}

\rev{For a nominal ET state~$(\etavec_0,\bm{\beta}_0)$ with a calibrated scattering profile, the beamforming design minimizes $\mathrm{CRB}_{\eta}(\etavec_0;\Rmat_x)$ for the center, orientation, and size parameters subject to user SINR and power constraints. }
\begin{align}
\underset{\{\wvec_k\},\bm W_0}{\text{minimize}}\quad
& \mathrm{CRB}_{\eta}(\etavec_0;\Rmat_x) \notag\\
\text{subject to}\quad
& \gamma_k \ge \Gamma_k,\quad k=1,\ldots,K, \notag\\
& \tr(\Rmat_x)\le P_{\max}, \notag\\
& \bm W_0 \succeq 0,
\label{eq:isac_opt}
\end{align}
where $\Gamma_k$ is the minimum SINR requirement for user~$k$. Here $\mathrm{CRB}_{\eta}(\etavec_0;\Rmat_x)=\tr(\Jmat(\etavec_0;\Rmat_x)^{-1})$. The ET parameters are fixed at the nominal point, and the transmit variables are optimized to minimize the joint CRB of position, orientation, and size. \rev{The main challenges are twofold: the CRB objective is nonlinear in the beamforming variables and coupled with the communication constraints through the shared transmit covariance~$\Rmat_x$, while the covariance-domain formulation leads to high-dimensional matrix variables and hence high computational complexity.}

% ============================================================
\section{Reduced-Dimensional Beamforming Design}\label{sec:beamdesign}

\begin{figure*}[!t]
  \centering
  \includegraphics[width=0.88\textwidth]{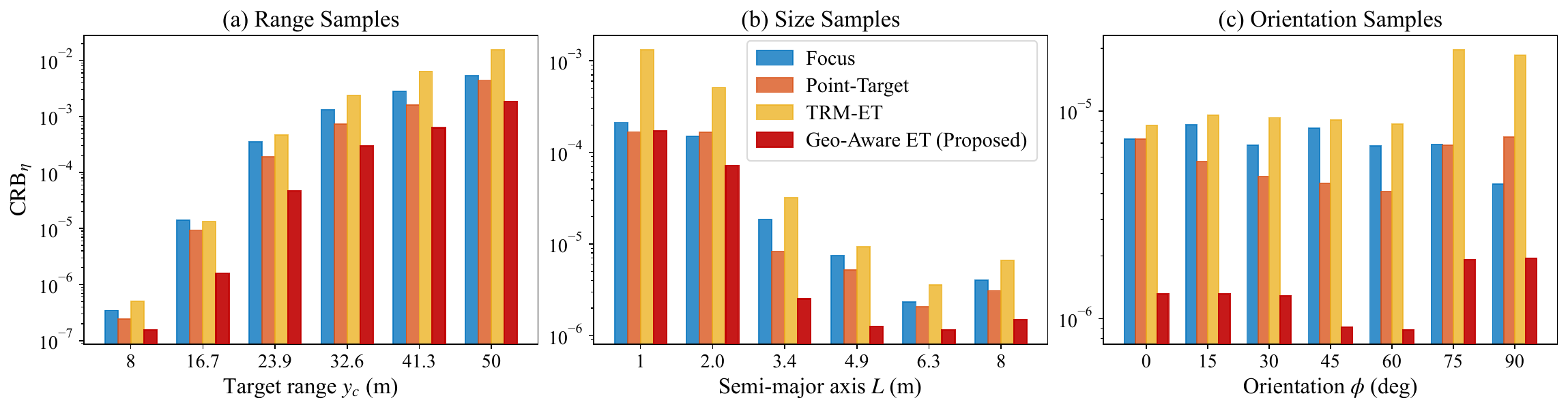}\vspace{-2pt}
  \caption{\rev{$\mathrm{CRB}_{\eta}$} at representative range, size, and orientation samples around the nominal ET state ($N\!=\!64$, $P_{\max}\!=\!0$~dBW).}
  \label{fig:size_distance}
\end{figure*}

\subsection{\rev{Reduced-Subspace Representation}}

\rev{For the covariance-domain analysis below, define $\bm W_k = \wvec_k \wvec_k^{\mathsf H} \succeq 0$ for $k=1,\ldots,K$, so that $\operatorname{rank}(\bm W_k)=1$, while $\bm W_0 \succeq 0$ already denotes the sensing covariance. Theorem~1 shows that the covariance variables of the lifted SDR can be restricted, without loss of relaxed optimality, to a low-dimensional subspace generated by the user channels and the ET steering/derivative directions.}

\begin{algorithm}[t]
\caption{Proposed ET-CRB Optimization}
\label{alg:et_sdp}
\textbf{Input:} $(\etavec_0,\bm{\beta}_0)$, $\{\bm h_k\}_{k=1}^{K}$, $\{\Gamma_k\}_{k=1}^{K}$, and $P_{\max}$\\
\textbf{Output:} ET-aware design
\begin{algorithmic}[1]
\STATE Build $\{\avec_m\}_{m=1}^{M}$ and $\{[\bar{\bm A}_m]_{:,p}\}_{m,p}$ at $\etavec_0$
\STATE \rev{Form the sensing matrices $\{\bm Q_{pq}\}$ via~\eqref{eq:q_def}}
\STATE \rev{Build the ET-aware subspace $\mathcal{U}$ from Theorem~1 and~\eqref{eq:set_et}--\eqref{eq:u_def}}
\STATE Obtain an orthonormal basis matrix $\bm U$ of $\mathcal{U}$
\STATE \rev{Parameterize $\bm W_0$ and $\bm W_k$ as in~\eqref{eq:reduced_param}}
\STATE \rev{Solve the reduced SDR~\eqref{eq:reduced_sdp}}
\STATE \rev{Recover the communication beams via~\eqref{eq:rankone_recovery}}
\end{algorithmic}
\end{algorithm}
\noindent\textbf{Theorem~1} \textit{(ET-aware reduced subspace).}
For the fixed nominal geometry~$\etavec_0$, define
\begin{equation}\label{eq:set_et}
  \mathcal{S}_{\mathrm{ET}} = \operatorname{span}\!\left(\{\avec_m\}_{m=1}^{M}, \{[\bar{\bm A}_m]_{:,p}\}_{m=1,\ldots,M;\,p=1,\ldots,D_\eta}\right)
\end{equation}
and
\begin{equation}\label{eq:u_def}
  \mathcal{U} = \operatorname{span}\!\big(\mathcal{R}(\bm H), \mathcal{S}_{\mathrm{ET}}\big), \qquad \bm H=[\bm h_1,\ldots,\bm h_K].
\end{equation}
\rev{Then there exists an optimal solution to the lifted SDR introduced below} such that
\begin{equation}
  \mathcal{R}(\bm W_k^\star) \subseteq \mathcal{U}, \qquad k=0,1,\ldots,K.
\end{equation}
Hence, if $\bm U$ is an orthonormal basis of~$\mathcal{U}$, each covariance variable of the lifted SDR can be written as $\bm W_k = \bm U\bm X_k\bm U^{\mathsf H}$ with $\bm X_k \succeq 0$, so the search over full-size $N\times N$ matrices is replaced by a search over reduced $r\times r$ matrices, where $r=\dim(\mathcal{U})\ll N$.

\begin{IEEEproof}
\rev{See Appendix~A.}
\end{IEEEproof}

\subsection{Semidefinite Relaxation and Reduced SDP}
\rev{By exploiting the reduced-subspace representation in \textbf{Theorem 1}, a equivalent reduced formulation of the original problem can be formulated. In particular,}
With $\Rmat_x = \sum_k \bm W_k + \bm W_0$, dropping the rank-one constraints $\bm W_k=\wvec_k\wvec_k^{\mathsf H}$ yields the SDR
\begin{align}
\underset{\{\bm W_k\},\bm W_0,\bm \Phi}{\text{minimize}}\quad
& \tr(\bm \Phi) \notag\\
\text{subject to}\quad
& \begin{bmatrix} \Jmat(\etavec_0;\Rmat_x) & \bm I_{D_\eta} \\ \bm I_{D_\eta} & \bm \Phi \end{bmatrix} \succeq 0, \notag\\
& \Gamma_k \Bigl(\sum_{j\ne k}\bm h_k^{\mathsf H}\bm W_j\bm h_k
 + \bm h_k^{\mathsf H}\bm W_0\bm h_k+\sigma_c^2\Bigr) \notag\\
& \qquad \le \bm h_k^{\mathsf H}\bm W_k\bm h_k,\quad \forall k, \notag\\
& \tr\!\Bigl(\sum_k \bm W_k+\bm W_0\Bigr)\le P_{\max}, \notag\\
& \bm W_k\succeq 0,\ \forall k,\qquad \bm W_0\succeq 0.
\label{eq:sdp}
\end{align}
Only the communication covariances are rank-constrained in the original formulation. Theorem~1 directly reduces the search space: instead of optimizing $(K+1)$ Hermitian matrices in $\C^{N\times N}$, the relaxed problem can be solved over $(K+1)$ reduced matrices in $\C^{r\times r}$, where $r=\dim(\mathcal{U})\ll N$. Accordingly, an optimal SDR solution can be sought within~$\mathcal{U}$, i.e.,
\begin{equation}\label{eq:reduced_param}
  \bm W_0=\bm U\bm X_0\bm U^{\mathsf H},\qquad
  \bm W_k=\bm U\bm X_k\bm U^{\mathsf H},\quad k=1,\ldots,K,
\end{equation}
where $\bm U\in\C^{N\times r}$ is an orthonormal basis of~$\mathcal{U}$ and $\bm X_k\in\C^{r\times r}$ satisfies $\bm X_k\succeq0$. Substituting into~\eqref{eq:sdp} gives the reduced SDR
\begin{align}
\underset{\{\bm X_k\},\bm \Phi}{\text{minimize}}\quad
& \tr(\bm \Phi) \notag\\
\text{subject to}\quad
& \begin{bmatrix} \Jmat(\etavec_0;\bm U\bar{\Rmat}_x\bm U^{\mathsf H}) & \bm I_{D_\eta} \\ \bm I_{D_\eta} & \bm \Phi \end{bmatrix} \succeq 0, \notag\\
& \Gamma_k \Bigl(\sum_{j\ne k}\bar{\bm h}_k^{\mathsf H}\bm X_j\bar{\bm h}_k
 + \bar{\bm h}_k^{\mathsf H}\bm X_0\bar{\bm h}_k+\sigma_c^2\Bigr) \notag\\
& \qquad \le \bar{\bm h}_k^{\mathsf H}\bm X_k\bar{\bm h}_k,\quad \forall k, \notag\\
& \tr\!\Bigl(\sum_{k=0}^{K}\bm X_k\Bigr)\le P_{\max}, \notag\\
& \bm X_k\succeq 0,\qquad k=0,1,\ldots,K,
\label{eq:reduced_sdp}
\end{align}
where $\bar{\Rmat}_x=\sum_{k=0}^{K}\bm X_k$ and $\bar{\bm h}_k=\bm U^{\mathsf H}\bm h_k$. Here $\bm X_0$ denotes the reduced sensing covariance, whereas $\{\bm X_k\}_{k=1}^{K}$ denote the reduced communication covariances. Thus, the ambient dimension~$N$ is replaced by the subspace dimension~$r$ without loss of the relaxed optimum.

Moreover, any feasible SDR point with $\bm h_k^{\mathsf H}\bm W_k\bm h_k>0$ for all~$k$ can be converted into another feasible point with rank-one communication covariances and the same objective value via
\begin{equation}\label{eq:rankone_recovery}
  \tilde{\bm w}_k=\frac{\bm W_k\bm h_k}{\sqrt{\bm h_k^{\mathsf H}\bm W_k\bm h_k}},\qquad
  \widetilde{\bm W}_0=\bm W_0+\sum_{k=1}^{K}\bigl(\bm W_k-\tilde{\bm w}_k\tilde{\bm w}_k^{\mathsf H}\bigr).
\end{equation}
Then $\tilde{\bm w}_k\tilde{\bm w}_k^{\mathsf H}\preceq\bm W_k$, the useful signal terms are preserved, and the total covariance $\Rmat_x$ remains unchanged, so the objective and all constraints are preserved. For the practically relevant case $\Gamma_k>0$, every feasible point satisfies $\bm h_k^{\mathsf H}\bm W_k\bm h_k>0$. Hence any optimal SDR solution is directly realizable. \rev{In other words, the SDR is tight and the relaxed optimum coincides with the original problem.} If $\mathcal{R}(\bm W_k)\subseteq\mathcal{U}$, then also $\tilde{\bm w}_k\in\mathcal{U}$.
% ============================================================
\section{Numerical Results}\label{sec:numerical}

All methods satisfy the same SINR and power constraints and are evaluated by the same $\mathrm{CRB}_{\eta}$, with $\Jmat$ assembled from the ET geometry and calibrated scattering profile of each tested scenario. For each sampled geometry in Fig.~\ref{fig:size_distance}, the ET steering and Jacobian matrices are rebuilt and all SDP-based methods are re-solved, so the geometry sweep compares per-scenario design quality rather than the robustness of one fixed covariance.

We compare Focus, Point-Target, TRM-ET, and Geo-Aware ET (Proposed). Focus uses MRT communication beams with the remaining power allocated to $\bm W_0\propto\avec(x_c,y_c)\avec(x_c,y_c)^{\mathsf H}$. Point-Target follows~\cite{liu2021crb_jrc}. TRM-ET adapts~\cite{liu2021crb_jrc} using $\bm G$ but without the parametric geometry. \rev{For numerical evaluation, the elliptical instance uses $\etavec=[x_c,y_c,\phi,L,b]^{\mathsf T}$ with $D_\eta=5$ and $M=76$ representative scattering points: $50$ on the outer ellipse, $25$ on a concentric inner ellipse scaled by $0.5$, and one center point. The outer-ellipse points are}
\begin{equation}
{\color{red}
\pvec_m(\etavec)=
\begin{bmatrix}
x_c\\ y_c
\end{bmatrix}
+
\bm R(\phi)
\begin{bmatrix}
L\cos\vartheta_m\\
b\sin\vartheta_m
\end{bmatrix}.}
\end{equation}
\rev{The rotation matrix is}
\begin{equation}
{\color{red}
\bm R(\phi)=
\begin{bmatrix}
\cos\phi & -\sin\phi\\
\sin\phi & \cos\phi
\end{bmatrix}.}
\end{equation}
\rev{The inner-ellipse points use $(0.5L,0.5b)$.}
\begin{table}[!t]
\caption{Main simulation parameters.}
\label{tab:sim_setup}
\centering
\footnotesize
\setlength{\tabcolsep}{3pt}
\begin{tabular}{@{}ll@{}}
\toprule
Parameter & Value \\
\midrule
Array & $N=64$ ULA, $f_c=4.9$~GHz \\
ET geometry & $(0,20)$~m, $\phi=30^\circ$, $L=3$~m, $b=0.8$~m \\
Users & $K=2$: $(15\,\text{m},-25^\circ)$ and $(18\,\text{m},35^\circ)$ \\
QoS / noise & $\Gamma=15$~dB, $\sigma_s^2=\sigma_c^2=-30$~dBm \\
Snapshots & $T=16$ \\
\bottomrule
\end{tabular}
\end{table}
Fig.~\ref{fig:size_distance} shows $\mathrm{CRB}_{\eta}$ at representative operating points selected from the underlying range, size, and orientation sweeps. Across the three views, the proposed design gives the smallest value of $\mathrm{CRB}_{\eta}$ at most operating points, with the clearest advantage at shorter ranges, moderately extended targets, and non-degenerate orientations. \rev{Around $y_c\approx 16.7$~m, the proposed design shows a pronounced advantage over Point-Target; around $L\approx 3.4$~m, it remains clearly below both Point-Target and TRM-ET.} The gap narrows for the smallest-size and farthest-range samples, where the ET response becomes closer to a point-target model.

\begin{figure}[!t]
  \centering
  \includegraphics[width=0.8\columnwidth]{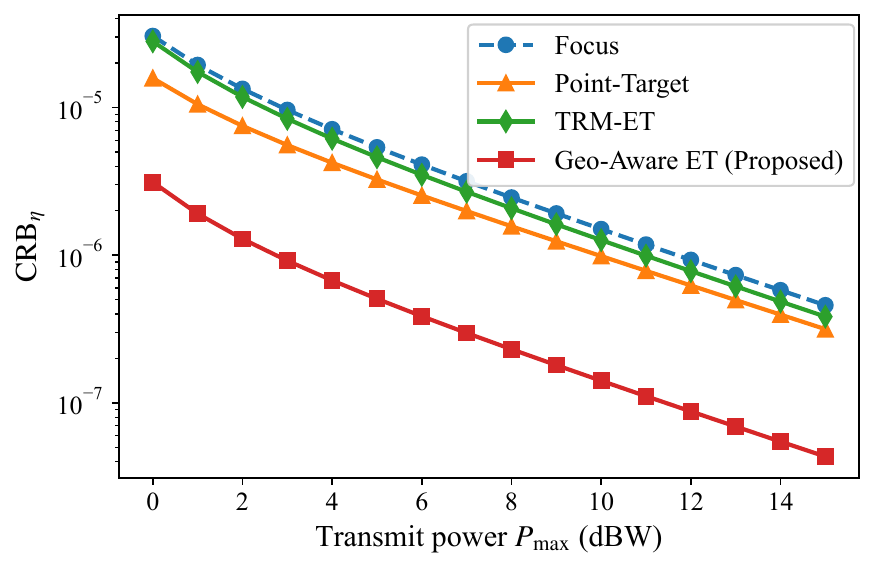}
  \caption{\rev{$\mathrm{CRB}_{\eta}$} versus transmit power ($L=5$~m, $b=1.5$~m).}
  \label{fig:crb_vs_power}
\end{figure}

Fig.~\ref{fig:crb_vs_power} plots $\mathrm{CRB}_{\eta}$ versus $P_{\max}$ for an enlarged target with $L=5$~m and $b=1.5$~m. All curves decrease with power, but the proposed design remains the best throughout the sweep. \rev{At $P_{\max}=0$~dBW, its $\mathrm{CRB}_{\eta}$ remains clearly below Point-Target and Focus, while TRM-ET stays between Point-Target and Focus.} The persistent gap at high power shows that the advantage is not confined to a low-SNR regime. This is expected because a larger ET provides more geometric diversity that a point-target model cannot exploit.

\begin{figure}[!t]
  \centering
  \begin{minipage}[t]{0.485\columnwidth}
    \centering
    \includegraphics[width=\textwidth]{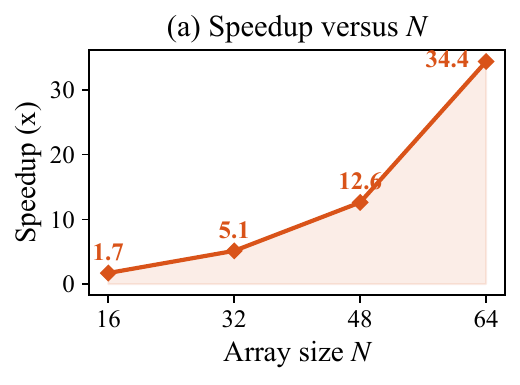}
  \end{minipage}\hfill
  \begin{minipage}[t]{0.485\columnwidth}
    \centering
    \includegraphics[width=\textwidth]{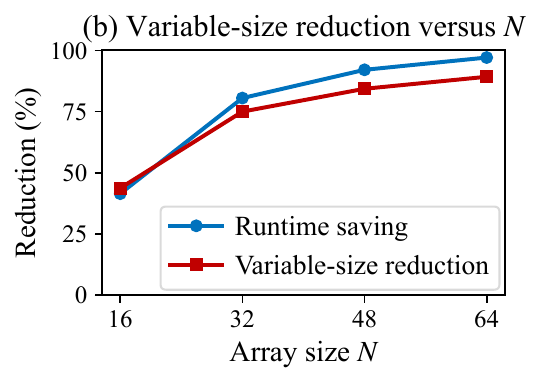}
  \end{minipage}
  \caption{Complexity gains of the reduced SDR as the array size increases.}
  \label{fig:sdp_complexity}
\end{figure}

Fig.~\ref{fig:sdp_complexity} shows that the advantage of the reduced formulation grows quickly with~$N$. \rev{The left panel shows that the runtime gain becomes increasingly pronounced as $N$ grows,} while the right panel shows that the variable-size reduction grows from about $44\%$ to about $89\%$. At $N=64$, the reduced subspace has $r=21$, so the matrix variable shrinks from $4096$ to $441$ entries. \rev{The small $r$ reflects the strong linear dependence among the ET steering and Jacobian vectors induced by the smooth contour and finite aperture.}

% ============================================================
\section{Conclusion}\label{sec:conclusion}

This letter studied communication-constrained beam design for a parametric near-field ET. A CRB formulation for the geometric parameters and an exact ET-aware reduced subspace \rev{led} to a reduced SDR together with a rank-one recovery procedure for the beamformer. The numerical results \rev{showed} consistent CRB gains over the baselines and large complexity savings for large arrays. Robust designs under scattering-profile mismatch and extensions to multi-target and wideband settings remain for future study.

% ============================================================
\appendices
\balance{}
\section{Proof of Theorem 1}

\begin{IEEEproof}
Let $\Pi_{\mathcal{U}}$ and $\Pi_{\mathcal{S}}$ denote the orthogonal projectors onto $\mathcal{U}$ and $\mathcal{S}_{\mathrm{ET}}$, respectively. For any feasible covariance tuple $\{\bm W_k\}_{k=0}^{K}$ of the lifted SDR~\eqref{eq:sdp}, define
\begin{equation}
  \widetilde{\bm W}_k=\Pi_{\mathcal{U}}\bm W_k\Pi_{\mathcal{U}},\qquad
  \widetilde{\Rmat}_x=\sum_{k=0}^{K}\widetilde{\bm W}_k=\Pi_{\mathcal{U}}\Rmat_x\Pi_{\mathcal{U}}.
\end{equation}
\noindent\textit{1) Feasibility is preserved:}
Since $\Pi_{\mathcal{U}}$ is Hermitian and idempotent, $\widetilde{\bm W}_k\succeq0$ whenever $\bm W_k\succeq0$, and
\begin{equation}
  \tr(\widetilde{\Rmat}_x)=\tr(\Pi_{\mathcal{U}}\Rmat_x)\le \tr(\Rmat_x),
\end{equation}
because $0\preceq\Pi_{\mathcal{U}}\preceq\bm I$. Hence the positive semi-definite (PSD) and power constraints are preserved. Moreover, for every user~$\ell$, $\bm h_\ell\in\mathcal{R}(\bm H)\subseteq\mathcal{U}$ implies $\Pi_{\mathcal{U}}\bm h_\ell=\bm h_\ell$, so
\begin{equation}
  \bm h_\ell^{\mathsf H}\widetilde{\bm W}_j\bm h_\ell
  =\bm h_\ell^{\mathsf H}\bm W_j\bm h_\ell,\qquad \forall\,\ell,j,
\end{equation}
and all SINR terms remain unchanged.

\noindent\textit{2) The sensing objective is preserved:}
Each summand of $\bm F_p$ has both left and right factors in $\mathcal{S}_{\mathrm{ET}}$, hence $\bm F_p=\Pi_{\mathcal{S}}\bm F_p\Pi_{\mathcal{S}}$. Therefore,
\begin{equation}
  \bm Q_{pq}=\Pi_{\mathcal{S}}\bm Q_{pq}\Pi_{\mathcal{S}}
  =\Pi_{\mathcal{U}}\bm Q_{pq}\Pi_{\mathcal{U}},\qquad \forall\,p,q,
\end{equation}
where $\mathcal{S}_{\mathrm{ET}}\subseteq\mathcal{U}$ is used in the second equality. By cyclicity of trace,
\begin{align}
  \tr(\bm Q_{pq}\widetilde{\Rmat}_x)
  &=\tr(\bm Q_{pq}\Pi_{\mathcal{U}}\Rmat_x\Pi_{\mathcal{U}}) \notag\\
  &=\tr(\Pi_{\mathcal{U}}\bm Q_{pq}\Pi_{\mathcal{U}}\Rmat_x)
   =\tr(\bm Q_{pq}\Rmat_x),
\end{align}
for all $p,q$. Hence every entry of $\Jmat(\etavec_0;\widetilde{\Rmat}_x)$ is identical to that of $\Jmat(\etavec_0;\Rmat_x)$, so the CRB objective is preserved.

\noindent\textit{3) Exactness:}
Thus every feasible point can be replaced by another feasible point with the same objective value and with $\mathcal{R}(\widetilde{\bm W}_k)\subseteq\mathcal{U}$ for all $k=0,1,\ldots,K$. An optimal solution therefore exists inside~$\mathcal{U}$. If $\bm U$ is any orthonormal basis of~$\mathcal{U}$, then $\mathcal{R}(\bm W_k)\subseteq\mathcal{U}$ holds if and only if $\bm W_k=\bm U\bm X_k\bm U^{\mathsf H}$ for some $\bm X_k\succeq0$, which proves the exact reduced parameterization.
\end{IEEEproof}

\def\IEEEbibitemsep{0pt plus 0.5pt}

\end{document}